\documentclass[journal=jacsat,manuscript=article]{achemso}
\usepackage{longtable}
\usepackage{multirow}
\usepackage{amsmath}
\usepackage{subfigure}
\usepackage[usenames,dvipsnames]{color}
\SectionNumbersOn
\author{Marcos H. Gim\'{e}nez}
\affiliation{Departamento de F\'{\i}sica Aplicada, Universitat
Polit\`{e}cnica de Val\`{e}ncia, Cam\'{\i} de Vera s/n, 46022,
Val\`{e}ncia, Spain.}

\author{Juan C. Castro-Palacio}
\affiliation{Department of Earth Sciences and Engineering, Faculty
of Engineering, Imperial College, London SW7 2AZ, United Kingdom.}

\author{Juan A. Monsoriu}
\email{jmonsori@fis.upv.es} \affiliation{Centro de Tecnolog\'{\i}as
F\'{\i}sicas, Universitat Polit\`{e}cnica de Val\`{e}ncia, Cam\'{\i}
de Vera s/n, 46022, Val\`{e}ncia, Spain.}
\date{\today}

\title{Direct visualization of mechanical beats by means of an oscillating smartphone}

\begin{document}

\noindent The resonance phenomenon is widely known from physics
courses\cite{hal:1999}. Qualitatively speaking, resonance takes
place in a driven oscillating system whenever the frequency
approaches the natural frequency, resulting in maximal oscillatory
amplitude.

\noindent
 Very closely related to resonance is the phenomenon of mechanical beating, which occurs when
the driving and natural frequencies of the system are slightly
different. The frequency of the beat is just the difference of the
natural and driving frequencies.

\noindent Beats are very familiar in acoustic systems. There are
several works in this journal on visualizing the beats in acoustic
systems\cite{kuhn:2014,gan:2015,keep:2015}. For instance, the
microphone and the speaker of two mobile devices were used in
previous work\cite{kuhn:2014} to analyze the acoustic beats produced
by two signals of close frequencies. The formation of beats can also
be visualized in mechanical systems, such as a mass-spring system
\cite{ande:2005} or a double-driven string \cite{car:2004}.

\noindent Here, the mechanical beats in a smartphone-spring system
are directly visualized in a simple way. The frequency of the beats
is measured by means of the acceleration sensor of a smartphone,
which hangs from a spring attached to a mechanical driver. This
laboratory experiment is suitable for both High School and first
year University physics courses.

\section{Equations governing the beats}
\begin{figure}[H]
\centering
\includegraphics[scale=0.50]{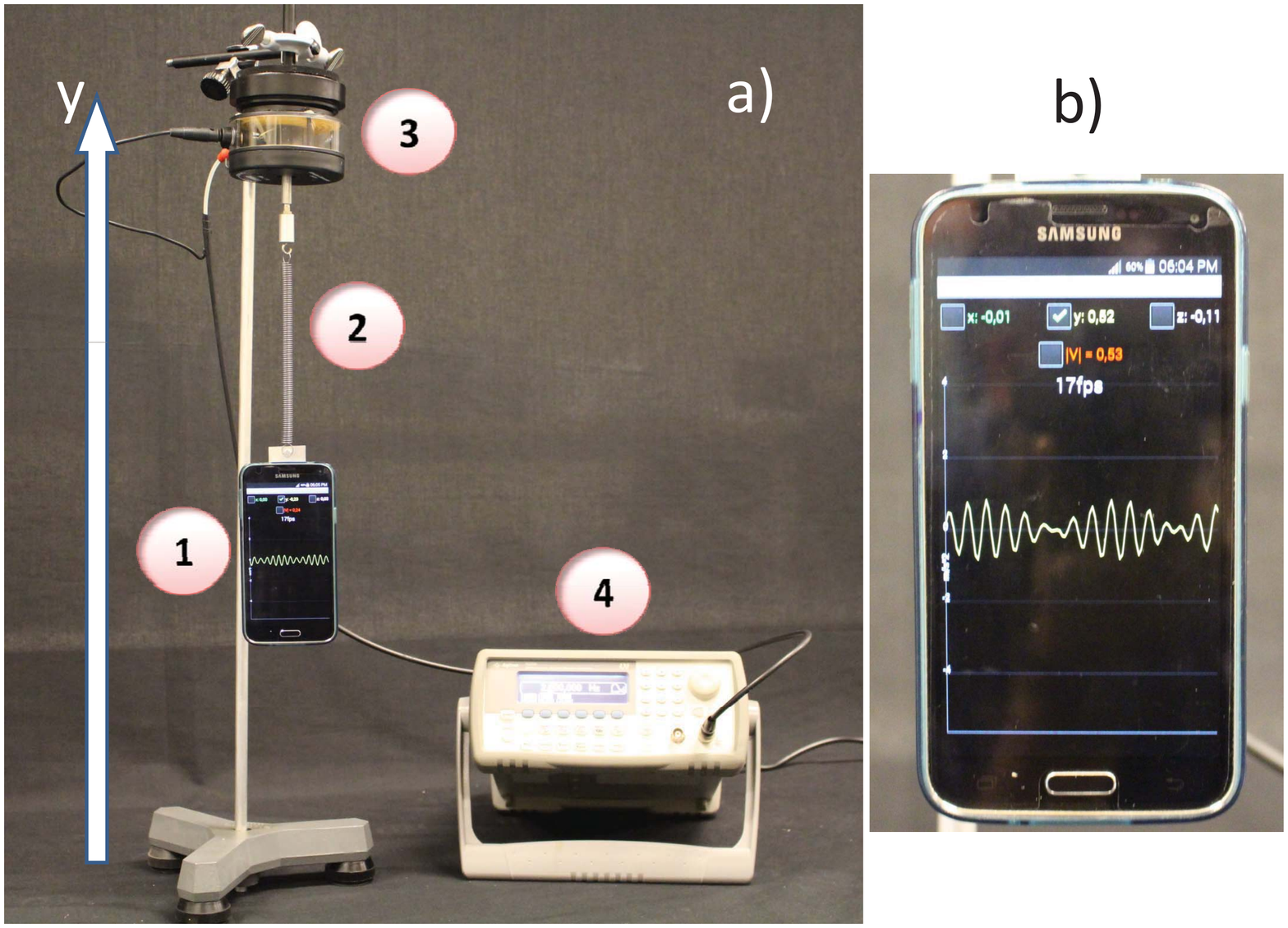}
\caption{The experimental setup consisting of the smartphone (1),
the spring (2), mechanical driver (3), and the AC generator (4) is
included in panel a. The smartphone has been attached to the spring
by means of a small metal plaque and a double-sided adhesive tape. A
smartphone showing beats on the screen is included on the right hand
side, in panel b. The y-axis is used as a reference axis.}
\label{fig1}
\end{figure}

From the experimental setup in Figure \ref{fig1}, the Second
Newton's Law can be stated as follows:

\begin{equation}
F_0sin(\Omega t+\phi)-ky-bv=ma, \label{newton}
\end{equation}

\noindent where $F_0sin(\Omega t+\phi)$ is the driving force acting
on the system, $F_0$ is its amplitude, $\Omega$ its frequency and
$\phi$ the initial phase. The term $-ky$ represents the elastic
force exerted by the spring on the smartphone and $k$, the spring
constant. The term $-bv$ is the damping force exerted by the air on
the system, where $b$ is the damping coefficient and $v$ the
velocity of the oscillating smartphone. On the right hand side of
the equation are the mass of the smartphone $m$, and its
acceleration $a$. The natural frequency of the system is
$\omega_0=\sqrt{\frac{k}{m}}$.

\noindent The solution of the above equation for a free oscillating
system, that is, without damping ($b=0$) and without driving forces
($F_0=0$), can be expressed as,

\begin{equation}
y_1(t)=Asin(\omega_0 t+\varphi).
\end{equation}

\noindent If the damping force is present, the solution is a little
more complicated and results in a damped oscillation,

\begin{equation}
y_1(t)=Ae^{-t/\tau}sin(\omega t+\varphi),
 \label{homo}
\end{equation}

\noindent where $\tau=2m/b$ is the relaxing time of the damped
oscillations and,
\begin{equation}
\omega=\sqrt{(\omega_0^2-(1/\tau)^2)}.
\end{equation}

\noindent Eq. \ref{homo} is the solution of the homogeneous equation
governing the damped oscillations (Eq. \ref{newton}). Moreover, we
can define a particular solution of Eq. \ref{newton}
as,\cite{gaf:2002}
\begin{equation}
y_2(t)=Dsin(\Omega t+\delta),
 \label{solution1}
\end{equation}

\noindent
where the amplitude is,
\begin{equation}
D=\frac{F_0}{m((\Omega^2-\omega_0^2)^2 + (2\Omega/\tau)^2)^{1/2}}.
 \label{final}
\end{equation}

\noindent
Finally, the general solution of Eq. \ref{newton} is then
given by the sum of both solutions given in Eq. \ref{homo} and
\ref{solution1},
\begin{equation}
y(t) = y_1(t)+y_2(t) = Ae^{-t/\tau}sin(\omega t+\varphi) +
Dsin(\Omega t+\delta)
 \label{solution3}
\end{equation}

\noindent A small damping force ($b\approx0$) would mean a large
$\tau$. Then, the experimental term becomes $e^{-t/\tau}\approx1$.
Under this assumption, the above equation reduces to,
\begin{equation}
y(t) = Asin(\omega_0 t+\varphi) + Dsin(\Omega t+\delta),
 \label{solution4}
\end{equation}
being $D=\frac{F_0}{m\left|\Omega^2-\omega_0^2\right|}$. The beats
appear in the system of Figure \ref{fig1} when the frequencies
$\Omega$ and $\omega_0$ in Eq. \ref{solution4} are only slightly
different. In our experimental setup, the oscillation data are
captured with the accelerometer of the smartphones. In this respect,
the expression for the acceleration of the system should be defined
from Eq. \ref{solution4} as,
\begin{equation}
a(t) = \frac{d^2y}{dt^2}=Bsin(\omega_0 t+\varphi) + Esin(\Omega
t+\delta),
 \label{solution5}
\end{equation}
where $B=-\omega_0^2A$ and $E=-\Omega^2D$. Using some trigonometric
identities, Eq. \ref{solution5} can be rewritten as
\begin{equation}
a(t) = C sin\left(\frac{\Omega+\omega_0}{2}t+\varphi_s\right),
 \label{solution6}
\end{equation}
with
$C=\sqrt{B^2+E^2+2BEcos(\left|\Omega-\omega_0\right|t+\varphi_c)}$.

\noindent The phases $\varphi_s$ and $\varphi_c$ can be derived from
the different parameters involved in the system, but this is not
relevant for the visualization of the beats in this work. It can be
noticed that the frequency in Eq. \ref{solution6} is the average
frequency, $(\Omega+\omega_0)/2$. On the other hand, the amplitude
$C$ is modulated by an envelope curve of frequency
$\left|\Omega-\omega_0\right|$.

\section{Description of the experiments and analysis}
\begin{figure}[H]
\centering
\includegraphics[scale=0.50]{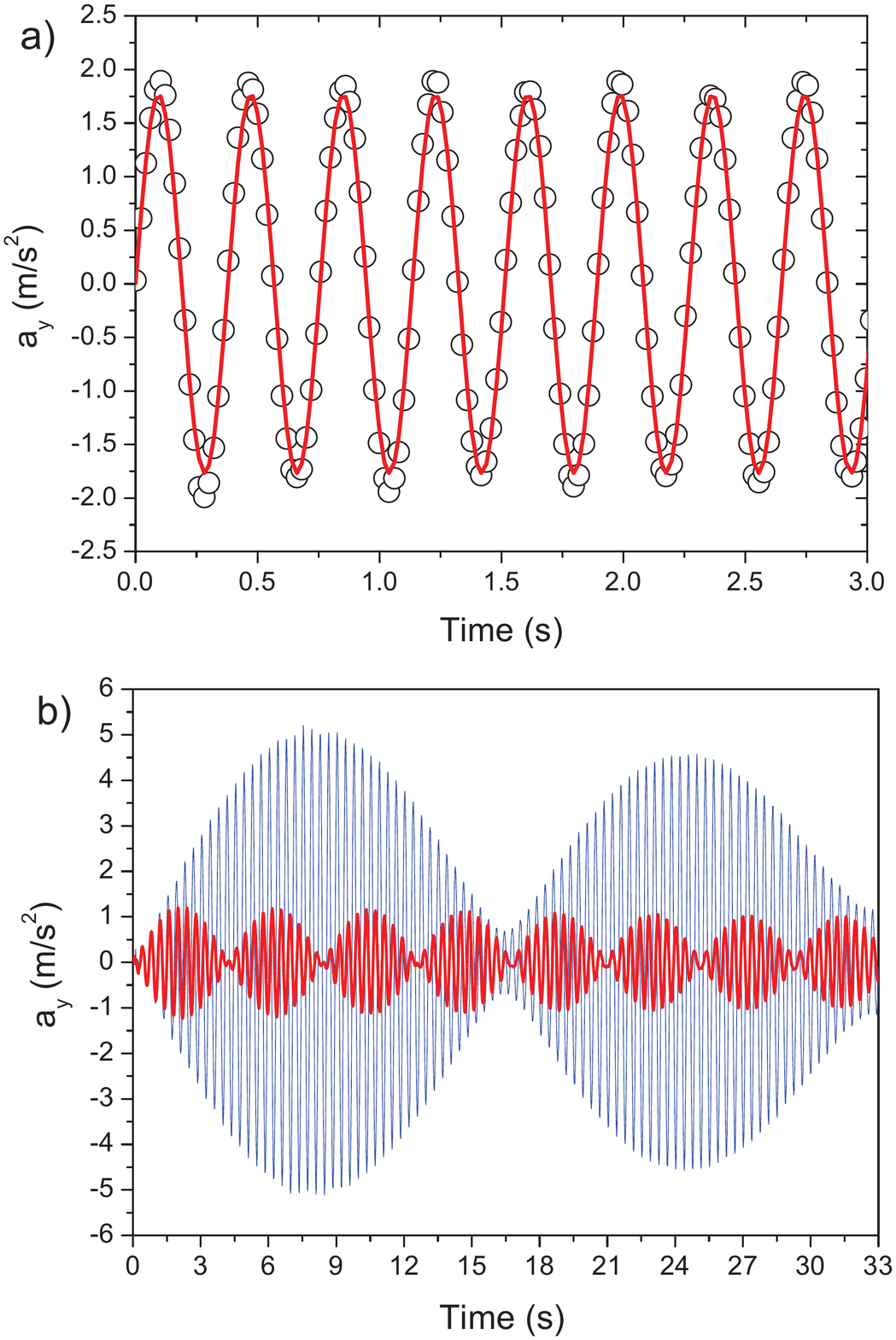}
\caption{Free oscillations of the smartphone as recorded by the
acceleration sensor (panel a) and driven oscillations for
$\left|f-f_0\right| \approx$ 0.059 Hz (blue line) and 0.242 Hz (red
line) (panel b).} \label{fig2}
\end{figure}

First of all, the mechanical driver is turned off and the natural
frequency of the system is determined. The nearly free oscillations
captured by the acceleration sensor of the smartphone are shown in
Figure \ref{fig2}a. The sensor data are collected with the Android
application ``Accelerometer Monitor" which can be downloaded from
Google Play website\cite{ac-mon}. A similar application called
``Vibsensor" is available for iOS.\cite{vib-sensor} The value of the
natural period, $T_0$, directly measured from the plot is
$T_0=0.369$ s, so the natural frequency without the driving force is
$f_0=1/T_0=2.710$ Hz. In the next experiments, the mechanical driver
is turned on. The driving frequency is set to $f=2.769$ (experiment
I) and $f=2.952$ Hz (experiment II), which corresponds to the
difference between the driving and the natural frequencies,
$\left|f-f_0\right|=\left|\Omega-\omega_0\right|/(2\pi)$, of 0.059
Hz and 0.242 Hz, respectively. The corresponding data registered
with the acceleration sensor for these frequencies are shown in
Figure \ref{fig2}b. The oscillation beats are clearly seen.

\noindent From direct measurements on Figure \ref{fig2}b, the values
of the beat frequencies ($f_{\rm beat}$) can be measured. Results
are included in Table \ref{tab1} along with the frequency difference
($\left|f-f_0\right|$) calculated from the value set at the AC
generator minus the natural frequency obtained from the fit (Figure
\ref{fig2}a) when the mechanical driver is off. Even when results
are not rigorously obtained, a good agreement is observed. That is
to say, the percentage deviations in the last column are lower than
$2\%$ for both cases.

\begin{table}[h]
  \caption{Driving frequency ($f$), frequency difference ($\left|f-f_0\right|$)
calculated from the value set at the AC generator minus the natural
frequency obtained from the fit (Figure \ref{fig2}a) when the
mechanical driver is off, and frequency of the beats ($f_{\rm
beat}$) measured on Figure \ref{fig2}b. The percentage deviations
($D$ $\%$) are included in the last column.}
\begin{center}
\begin{tabular}
[c]{ccccc}\hline\hline & Driving frequency  & Theoretical beat
frequency  & Experimental beat frequency  & $D$
\\
& $f$ (Hz) &  $\left|f-f_0\right|$ (Hz) & $f_{beat}$ (Hz) & ($\%$)\\
\hline
Exp. I & 2.769 &  0.059 & 0.060 & 1.7\\
Exp. II & 2.952 &  0.242  & 0.239 & 1.3\\\hline
\end{tabular}
\end{center}
\label{tab1}
\end{table}

\noindent This laboratory experiment is being implemented with
success in the first physics course for the Industrial Design
Engineering degree at the Universitat Polit\`{e}cnica de
Val\`{e}ncia, Spain. We carried out a survey which indicated that
over 95\% of the students bear a smartphone. The use of these
devices, which are very familiar to students, contributes greatly to
motivate introductory and first-year university students to physics
concepts.

\section*{Acknowledgments}
Authors would like to thank the Institute of Educational Sciences of
the Universitat Polit\`{e}cnica de Val\`{e}ncia (Spain) for the
support of the Teaching Innovation Groups MoMa and e-MACAFI and for
the financial support through the Project PIME 2015 B18.


\end{document}